\documentclass[prl,twocolumn,showpacs]{revtex4}
\usepackage{amsmath,amsfonts,amssymb}
\usepackage{bm,times}
\usepackage{graphicx}

\newcommand{\be}{\begin{equation}}
\newcommand{\ee}{\end{equation}}
\newcommand{\bea}{\begin{eqnarray}}
\newcommand{\eea}{\end{eqnarray}}
\newcommand{\ek}{\epsilon_{\mathbf{k}}}

\newcommand{\Ek}{E_{\mathbf{k}}}

\newcommand{\sumk}{\sum_{\mathbf{k}}}

\newcommand{\Omegaq}{\Omega_{\mathbf{q}}}

\newcommand{\uk}{u_{\mathbf{k}}}
\newcommand{\vk}{v_{\mathbf{k}}}

\begin{document}

\title{Thermodynamics of Interacting Fermions in Atomic Traps}

\author{Qijin Chen}
\author{Jelena Stajic}
\altaffiliation{Present address: Los Alamos Nat'l Lab,
Los Alamos, NM 87545}
\author{K. Levin}

\affiliation{James Franck Institute and Department of Physics,
University of Chicago, Chicago, Illinois 60637}

\date{\today}

\begin{abstract}
  We calculate the entropy in a trapped, resonantly interacting Fermi
  gas as a function of temperature for a wide range of magnetic fields
  between the BCS and Bose-Einstein condensation endpoints. This
  provides a basis for the important technique of
  adiabatic sweep thermometry, and serves to
  characterize quantitatively the evolution and nature of the
  excitations of the gas.  The results are then used to calibrate the
  temperature in several ground breaking experiments on $^6$Li and
  $^{40}$K.
  
\end{abstract}

\pacs{03.75.Hh, 03.75.Ss, 74.20.-z \hfill \textsf{Phys. Rev. Lett. \textbf{95},
  260405 (2005)}}

\maketitle


The claims \cite{Jin4,Ketterle3a,Thomas2a,Grimm3a,KetterleV} that
superfluidity has been observed in fermionic atomic gases have generated
great excitement. Varying a magnetic field, one effects a smooth evolution
from BCS superfluidity to Bose-Einstein condensation (BEC)
\cite{Eagles,Leggett}.
In this Letter we use a BCS-BEC crossover theory to study the entropy
$S$ over the entire experimentally accessible crossover regime.  Our
goal is to help establish a methodology for obtaining the temperature
$T$ of a strongly interacting Fermi gas via adiabatic sweeps. This
addresses an essential need of the experimental cold atom community by
providing a temperature calibration for their ground breaking
experiments \cite{Jin3,Grimm4a,Ketterle3a}.  In the process, we
characterize quantitatively the evolution of the excitations and show
how their character evolves smoothly from fermionic to bosonic, and
conversely.

In adiabatic sweeps, the starting $T$ at either a BEC or BCS endpoint is
estimated from the ``known" shape of the profile in the trapped cloud.
Then, the temperature (near unitarity, say) is obtained by equating the
entropy before the sweep to that in the strongly interacting regime
after the sweep.  Conventionally, the temperature scale which appears in
the superfluid phase diagram \cite{Jin3,Ketterle3a} involves an
isentropic sweep between the unitary and the non-interacting Fermi gas
regimes. The direction of the sweep is irrelevant in these reversible
processes.
The important experimental phase diagrams plot the condensate fraction,
$N_s/N$ near unitarity vs this Fermi gas-projected temperature,
$T_{\text{eff}}$.

In this paper, our thermodynamical calculations are used to relate the
actual physical temperatures $T$ to $T_{\text{eff}}$, where, in general,
$T$ is significantly greater than $T_{\text{eff}}$.  A calculation of
$N_s(T)$ is simultaneously undertaken \cite{footnoteonN0,ourreview}
which provides an important self-consistency condition on the
thermodynamics, since the same excitations appear in both, Moreover, a
calculation of $N_s$ has to be done with proper attention paid to
collective modes and gauge invariance \cite{Kosztin2}.  Here we address
the various condensate fractions found experimentally
\cite{Jin4,Ketterle3a}, (with emphasis on $^6$Li), as a function of
$T_{\text{eff}}$, in the experimental range of magnetic fields.

Our work is based on the BCS-Leggett ground state \cite{Eagles,Leggett}
and its finite $T$ extension \cite{ourreview}.  Four different classes
of experiments have been successfully addressed in this framework. These
include (i) $T \approx 0$ breathing modes experiments
\cite{Thomas2a,Grimm3a} and theory \cite{Tosia,Heiselberg}, (ii) radio
frequency (RF) pairing gap experiments \cite{Grimm4a} and theory
\cite{Torma2,yanhe}, and (iii) $T$-dependent density profiles
\cite{JS5}. Finally, (iv) plots of the energy $E$ vs $T$ at unitarity
\cite{ThermoScience} yield very good agreement with experiment and serve
to calibrate the present thermometry.  Two well-known weaknesses of the
mean field approach (an underestimate of $\beta$ and an overestimate of
the inter-boson scattering length $a_B$ in the deep BEC regime) should
be noted. The first affects $E(T)$ but not $S(T)$.  However, for the
second we introduce a caveat: if the initial endpoint of sweep
thermometry is sufficiently deep in the BEC regime (say, $k_F a \leq
0.3$) the accuracy of the final temperature, we calculate for the
unitary regime, could be improved by computing the initial $S$ in the
deep BEC regime using a pure-boson model with $a_B$ set by hand to the
Petrov result \cite{Petrov}.

Because previous thermodynamic theories did not address unitarity, it
has not been possible until now to arrive at a temperature scale in the
experimentally interesting resonant superfluid regime.
Carr \textit{et al.} \cite{Carr,Carr3} calculated $S$ at the BCS and
weakly interacting, deep-BEC endpoints.  The latter true Bose limit
which they considered does not appear to be appropriate to current
collective mode experiments, \cite{Thomas2a,Grimm3a}, which show
\cite{Tosia,Heiselberg} that for physically accessible (i.e., near-BEC)
fields, fermions are playing an important role. Thus, the BCS-Leggett
ground state appears to be more appropriate than one deriving from
Bose-liquid-based theory.
Williams \textit{et al.} \cite{Williams} calculated $S$ for a BCS-BEC
crossover theory using a mixture of noninteracting fermions and bosons
\cite{Williams}.  This work, omits the important and self consistently
determined fermionic excitation gap $\Delta$ which is an essential
component for describing the thermodynamics of fermionic superfluids.

Our thermodynamical calculations focus on this self-consistently
determined $\Delta$; they are based, for completeness, on a two-channel
Hamiltonian \cite{ourreview,Griffin,Milstein}.  Here $\Delta$ appears in
the fermionic dispersion $\Ek = \sqrt{(\ek-\mu)^2 + \Delta^2}$.  (We
define $\ek=\hbar^2 k^2/2m$ as the kinetic energy of free atoms, and
$\mu$ the fermionic chemical potential.)  Importantly, this $\Delta$
provides a measure of bosonic degrees of freedom. In the fermionic
regime ($\mu > 0)$, $\Delta$ is just the energy required to dissociate
the pairs and thereby excite fermions.  At finite $T$, the
closed-channel molecular bosons and the open-channel finite momentum
Cooper pairs are strongly hybridized with each other, making up the
``bosonic" excitations which contribute to thermodynamics.

Our many-body formalism has been described below the superfluid
transition temperature $T_c$\cite{ourreview}.  The parameter $\Delta$, (when
squared), is the analogue of the total number of particles, in the
simplest theory of BEC.  Just as in BEC, there are two self-consistency
conditions: (i) the effective chemical potential of the pairs,
$\mu_{pair}$, is zero, for $ T \leq T_c$ (as is that of the
closed-channel molecular bosons $\mu_{mb}$), and, (ii) the number of
pairs, reflected in $\Delta^2(T)$ contains two additive contributions
representing condensed ($\tilde{\Delta}_{sc}^2$) and noncondensed
($\Delta_{pg}^2$) pairs.  The first condition implies that $\Delta(T)$
satisfies a BCS-like gap equation.  Then, the condensate is deduced,
just as in BEC, by determining the difference between $\Delta^2$ and
$\Delta_{pg}^2$.  In this approach the hybridized pairs have dispersion
$\Omegaq = \hbar^2 q^2/2M^*$, with effective pair mass $M^*$.

We now extend this approach above $T_c$.  Our first equation represents
the important defining condition on $\mu_{pair}$: that the inverse pair
propagator (or $T$-matrix) $\left.  t^{-1}(Q)\right|_{Q \equiv 0} = Z
\mu_{pair}$, with (inverse) ``residue" $Z$.  While in the superfluid
regions $\mu_{pair}$ and $\mu_{mb}$ vanish, in general, we have
\begin{equation}
U^{-1}_{eff}(0) + \sum_{\bf k}
\frac{1-2 f(\Ek)}{2 \Ek}= Z\mu_{pair}  \,,
\label{eq:1}
\end{equation}
where $U_{eff}(0)=U+g^2/(2\mu-\nu)$ involves the sum of the direct
attraction $U$ between open-channel fermions, as well as the virtual
processes associated with the Feshbach resonance.  Here $f(x)$ is the
Fermi distribution function. The determination of the inter-channel
coupling constant, $g$, and the magnetic field detuning, $\nu$, is
described elsewhere \cite{ClosedChannel}, as are the residues $Z$ and
$Z_b$ \cite{ourreview}.  The contribution from hybridized bosons will
lead to a normal state excitation gap \cite{JS2a,ourreview,Grimm4a,Jin5}
or pseudogap (pg). This can be written in terms of the Bose distribution
function $b(x)$ as
\begin{equation}
\Delta_{pg}^2=Z^{-1} \sum_{\bf q}\, b(\Omega_q -\mu_{pair})\,.
\label{eq:2}
\end{equation}

We use the local density approximation (LDA) throughout with a
harmonic trap potential $V(r)$.  For notational simplicity, we omit
writing $V(r)$ in favor of $\mu(r)$ according to the LDA prescription:
$ \mu \rightarrow \mu(r) \equiv \mu -V(r)$, where $\mu \equiv \mu(0)$.
The total atomic number $N\equiv \int \mathrm{d}^3 r\, n(r)$, where
\begin{eqnarray}
n&=& 2 n_{b0} + 2Z_b^{-1} \sum_{\bf q} b(\Omega_q  -\mu_{mb})\nonumber\\
&&{}+ 2 \sum_{\bf k}\left [\vk^2 (1-f(\Ek))+\uk^2 f(\Ek)\right] \,.
\label{eq:3}
\end{eqnarray}
Here $n_{b0} = g^2 \Delta_{sc}^2/[(\nu -2 \mu(r))U]^2$ is the density
of condensed closed-channel molecules,
and $\uk^2, \vk^2 = [1\pm (\ek -\mu(r))/\Ek]/2$.  The total
order parameter \cite{Griffin,Milstein,JS2a} is given by
$\tilde{\Delta}_{sc} = \Delta_{sc} + |g|\sqrt{n_{b0}}$.

\begin{figure}
\includegraphics[width=2.5in,clip]{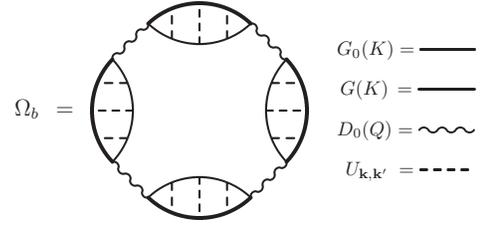}
\caption{Bosonic contribution to the thermodynamical potential. Here
$G_0$ ($G$) and $D_0$ ($D$) are the ``bare'' (``full'') propagators
associated with the fermions and closed-channel molecular bosons,
respectively, $K$ and $Q$ are four-momenta, and
$U_{\mathbf{k},\mathbf{k'}}$ is the open-channel pairing interaction.
}
\label{fig:1}
\end{figure}

To make progress, we numerically solve Eqs.~(\ref{eq:1})-(\ref{eq:2})
at each $r$ for given $\mu$ and then self-consistently adjust $\mu$
via the total number constraint.
Next we obtain the entropy $S$ directly from the thermodynamical
potential \cite{Chen4}.  This potential contains fermionic contributions
from bare fermions, $\Omega_f$, and bosonic contributions $\Omega_b$.
The latter is given by the sum of all possible ring diagrams shown in
Fig.~\ref{fig:1}. It can be easily shown that this $\Omega_b$ is
consistent with the self energy diagrams for the fermions and the
molecular bosons.
After regrouping, we see that $S$ has two contributions, from fully
dressed fermions ($S_f$) and from their bosonic counterpart ($S_b$).
The total entropy involves an integral over the trap, given by $S=\int
\mathrm{d}^3r\, s(r)$ (and similarly for $S_f$ and $S_b$), where
\begin{eqnarray}
s &=&s_f + s_{b} \nonumber\,, \\
s_f &=& -2 \sumk [ f_k \ln f_k + (1-f_k) \ln (1-f_k)], \nonumber \\
s_{b} &=& - \sum_{q\ne 0} [b_q \ln b_q -(1+b_q) \ln (1+b_q) ], 
\end{eqnarray}
where $f_k \equiv f(\Ek)$, and $b_q \equiv b(\Omegaq-\mu_{mb})$; a
relatively small contribution associated with the $T$ dependence of
$\Omegaq$ has been dropped.  The fermion contribution coincides
formally with the standard BCS result for noninteracting
quasiparticles [although here $\Delta(T_c) \neq 0$].  And the bosonic
contribution is given by the expression for non-directly-interacting
bosons with dispersion $\Omegaq$.  These bosons are not free, however;
because of interactions with the fermions, their propagator contains
important self-energy and mass renormalization effects.

\begin{figure}
\includegraphics[width=3.0in,clip]{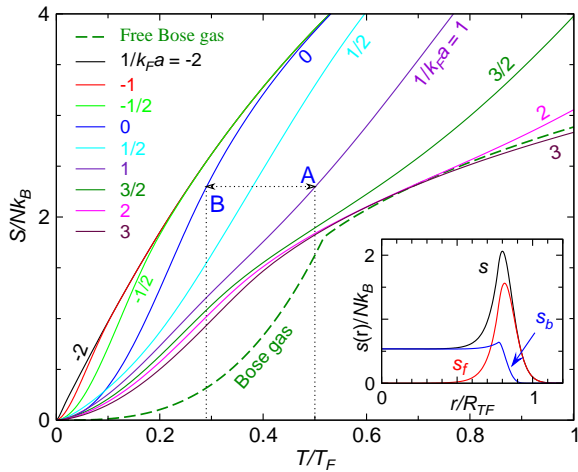}
\caption{(color online) Entropy per atom as a function of $T$ for
  different values of $1/k_Fa$ from BCS to BEC in a harmonic trap. 
  The dotted lines show an isentropic sweep between $1/k_Fa=1$ and
  unitarity. For comparison, we also plot $S$ for an ideal Bose gas
  (dashed line). The $1/k_Fa=3$ curve lies below the dashed line at
  $T>0.8T_F$ reflects that $M^*\ne 2m$.  The inset plots the spatial
  profile of total entropy $s$ (black curve) and its fermionic ($s_f$,
  red) and bosonic ($s_b$, blue) component contributions at unitarity
  for $T=T_c/4$.  Here $R_{TF}$ is the Thomas-Fermi radius, and
  $T_c =0.27T_F$.  }
\label{fig:2}
\end{figure}

Figure \ref{fig:2} illustrates the behavior of $S$ as a function of $T$
obtained from our self-consistent equations, over the entire
experimentally relevant crossover regime.  The magnetic field is
contained in the parameter $1/k_Fa$, which increases with decreasing
field.  Here $a$ is the $s$-wave fermionic scattering length, $k_F$ is
the Fermi wavevector at the trap center, and $k_BT_F=\hbar^2 k_F^2/2m$
is the noninteracting Fermi temperature.
Two important aspects of the fermionic contribution $S_f$ should be
noted.  Generally, the fermions have a gap $\Delta$ in their excitation
spectrum (which increases with decreasing field) and moreover this $T$
dependent gap is inhomogeneous so that the fermions near the trap edge
often behave as free particles at $T>\Delta$.  These quasi-``free''
fermions change the $T$ dependence of $S_f$ from exponential to power
law.  They have also been seen in RF experiments \cite{Grimm4a,Torma2},
as a free fermion peak in the spectra.

We refer to Fig.~\ref{fig:2}, starting from the high field or BCS regime
where $S$ is linear in $T$.  As the field is lowered towards unitarity,
$S_f$ will vary as a low-$T$ power law which is higher than linear.
Simultaneously, the bosonic degrees of freedom emerge.  Here one sees a
$T^{3/2}$ power law from these excited bosons. At unitarity, bosonic
effects dominate for $T/T_F \lesssim 0.05$ or $T/T_c \lesssim 0.2$. For
an extended range of $T<T_c$, the fermions and bosons combine to yield,
$S \propto T^2$, which can be compared with the experimental power law
\cite{ThermoScience} $T^{2.73}$.  Finally in the near-BEC regime one
sees an essentially pure bosonic $T^{3/2}$ power law in $S$ at low $T$
in the superfluid phase.
The relative contribution of the bosonic excitations, $S_b/S$ evolves
continuously from 0 to 1 as $1/k_Fa$ increases from $-\infty$ to
$+\infty$.  $S$ becomes dominantly bosonic once $\mu$ becomes negative.

The bosonic $T^{3/2}$ power law found in the trap is the same as found
for the homogeneous situation.  Inhomogeneity effectively disappears
here because the fermion-boson interactions lead to the self-consistent
constraint that $\mu_{pair}=0$ for the entire superfluid region.  This
same disappearance of inhomogeneity is found in Ref.~\cite{Carr}.  This
is different from a strictly non-interacting Bose system \cite{Williams}
(dashed line in Fig.~\ref{fig:2}) where one does not have a vanishing
boson chemical potential below $T_c$ except at the trap center.
The previous work of Ref.~\cite{Carr} is based on interacting
but true bosons. The present situation is more complex since 
Cooper pair operators do not obey Bose commutation relations,
(nor does the linear combination of Cooper pair and
closed-channel boson operators). This suggests that a theory
based on a true Bose liquid may not be appropriate for the
fields that have been accessed experimentally.
Moreover, if one were to contemplate contributions from the linearly
dispersing Goldstone bosons, albeit within a more general ground state,
their contribution, at unitarity, will not be as important as that from
the edge fermions.

To shed additional light on the component fermionic and bosonic
contributions, in the inset to Fig.~\ref{fig:2} we decompose the
various terms in the entropy to reveal their spatial distributions,
for the unitary case at $ T = T_c/4$.  It can be seen that the
fermionic contribution $s_f$ (red curve) is limited to the trap edge,
where $\Delta$ is small.  By contrast, the bosonic contribution $s_b$
(blue curve) is evenly distributed over the superfluid region and
rapidly decays at larger radii.

Figure \ref{fig:2} provides a basis for thermometry in adiabatic sweep
experiments.  The vertical lines illustrate how to choose an initial
temperature ($T_i = 0.5 T_F$ at point ``A'') with an initial value of
$1/k_Fa$ (=1) and use an isentropic sweep (represented by the
horizontal line through ``A'' and ``B'') to obtain the final
temperature ($T_f = 0.28 T_F$ at point ``B'') with the final value of
$1/k_Fa$ (=0).  It is most convenient to begin with either the BCS
regime or BEC regimes, since here $T_i$ can, in principle, be
determined by fitting the density profiles.

\begin{figure}
\includegraphics[width=3.2in,clip]{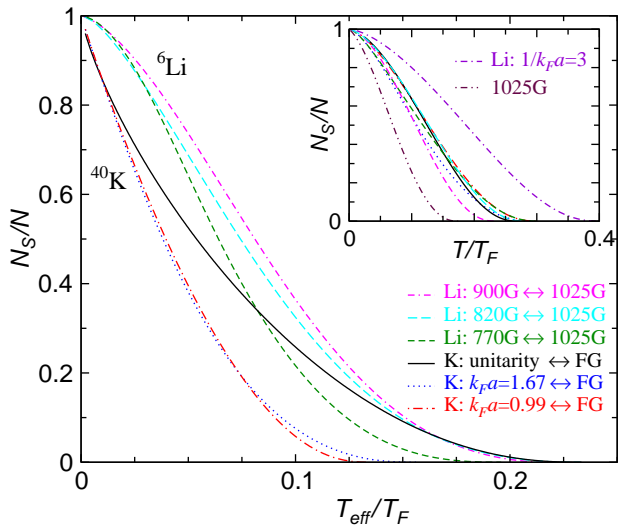}
\caption{(color online) Superfluid density $N_s/N$ at different magnetic
  fields for $^6$Li and $^{40}$K as a function of the effective
  temperature, $T_{\text{eff}}$, measured in a near-BCS (at 1025G for
  $^6$Li) or noninteracting Fermi gas (FG, $^{40}$K) state accessed via
  reversible adiabatic sweeps of magnetic field.  The inset plots the
  same $N_s/N$ as a function of the physical temperature at each field
  value. The system is not far from the resonance in these states.  Also
  plotted in the inset is $N_s(T)/N$ at 1025G and $1/k_Fa=3$ for $^6$Li.
  The values of field and $k_Fa$ were chosen based on
  Refs.~\cite{Ketterle3a} and \cite{Jin3}.  For $^6$Li, $T_F=3.6\mu$K,
  so that $1/k_Fa= -0.59, -0.26, 0.07, 0.38 $ for 1025~G, 900~G, 820~G
  and 770~G, respectively. }
\label{fig:4}
\end{figure}

Figure ~\ref{fig:4} presents a plot of the superfluid fraction
\cite{ourreview,footnoteonN0} $N_s/N$ in the intermediate regime as a
function of an effective temperature $T_{\text{eff}}/T_F$ for different
values of initial fields or $1/k_Fa$. Here $T_{\text{eff}}$ is the
temperature reached after an adiabatic sweep to a BCS-like state.  Based
on experiment, we take the final state as $1/k_Fa=-0.59$ at 1025~G for
$^6$Li \cite{Ketterle3a} and a noninteracting Fermi gas for $^{40}$K
\cite{Jin3}.

The same vertical axis appears in the inset but with the physical
temperature scale $T$, so that this figure provides a means of
directly calibrating the temperature scale $T_{\text{eff}}$ which has
been used in the important phase diagrams of $^6$Li and $^{40}$K.
Figure ~\ref{fig:4} also provides a means of comparing the condensate
fractions with those in the phase diagrams.  For $^6$Li at 900~G, with
$T_{\text{eff}}/T_F$ = 0.2, 0.1 and 0.05, the experimental condensate
fractions are 0.0, 0.1 and 0.6.  This should be compared with our
calculated values, 0.006 0.36, and 0.73, respectively.  For $^6$Li at
770~G, the condensate first appears at $T_{\text{eff}}/T_F= 0.18$,
consistent with theory.  From the values of $k_Fa$ and $T_c$ at both
ends (importantly, the latter can be read from the inset), one can
easily see that the sweep from 770~G to 1025~G is still very far from
a full BEC-BCS sweep.

There are two reasons for the larger condensate fractions found
theoretically for $^6$Li. A calculation of the BEC-like density profiles
shows that the noncondensed pairs inside the superfluid region have a
flat density distribution, which reflects the vanishing of $\mu_{pair}$
\cite{JS5}. The superfluid fraction extracted experimentally (assuming a
Gaussian form for the noncondensed particles inside the condensate core)
is, thus, underestimated, most notably around $T_c/2$.  In addition,
earlier work \cite{ThermoScience} shows that when the system is treated
as a noninteracting Fermi gas, $T_{\text{eff}}$ will be underestimated
whenever a condensate is present (at $ T < T_c \approx 0.17 T_F$ at
1025~G).  This suggests that theory and experiment can be brought into
rather good agreement for the case of $^6$Li.  For $^{40}$K, one has to
appeal to non-adiabaticity and other complications of the sweep process
to understand the small measured fractions.

For this case, we emphasize temperature scales.  For a full BCS to
near-BEC [$(k_Fa)_{\text{final}} = 1.7$] adiabatic sweep \cite{Jin3}
with initial $T_{\text{eff}}\equiv T_i = 0.19T_F$, the final reported
temperatures in experiment \cite{Jin3} and in theory are $T_f = 0.47
T_F$ and $0.33 T_F$, respectively.  For the same $(k_F
a)_{\text{final}}$ but with $T_i = 0.17T_F$, we find $T_f \approx T_c$,
in agreement with the observed sudden onset of a bimodal distribution in
the density profile.  Similarly for a sweep from a Fermi gas down to
$(k_F a)_{\text{final}} = 0.99$ with $T_i = 0.06 T_F$, the
experimentally quoted and theoretically calculated $T_f$ are $0.25T_F$
and $0.18T_F $, respectively.  The experimental sweeps were not strictly
adiabatic \cite{Jin3}, so that the experimental $T_f$ should serve as
upper bounds.  Our calculations are more consistent with experiment,
than if one had presumed a $T^3$ power law for $S$ in the BEC regime,
from which one would infer $T_f = 0.52T_F$ and $0.37T_F$, respectively,
exceeding the upper bounds.

Experimentally, $^{40}$K gases \cite{Jin3} are prepared in the
noninteracting limit, where, as a result of heating associated with an
adiabatic sweep, low $T$ is difficult to reach.  By contrast $^6$Li
gases \cite{Grimm3a,Ketterle3a} are prepared in the BEC regime, so that
higher condensate fractions of 80\% and 95\% have been reported
\cite{Grimm3a,Grimm4a,Ketterle3a} near unitarity via adiabatic cooling.
Finally, we note that the rather large $T_c \approx 0.17 T_F$ at 1025~G
makes it hard to access the Fermi gas regime in $^6$Li.  This may be
circumvented either by reduction in the size of $N$ or $T_F$ or,
possibly, by sweeps to 528~G \cite{Thomas}.  In $^{40}$K, one avoids
this problem altogether.

Without knowing the temperature, measurements in this field cannot be
directly compared to any theory.  The present work presents a theory for
the entropy $S$ of a Fermi gas, at general accessible magnetic fields,
which thereby calibrates $T$ in various existing
\cite{Jin3,Grimm4a,Ketterle3a} and future experiments.

We are extremely grateful to J.E. Thomas, J. Kinast and A. Turlapov
for many helpful discussions, and to N. Nygaard, C.  Chin, M. Greiner,
C.  Regal, D.S. Jin, and M. Zwierlein as well. This work was supported
by NSF-MRSEC Grant No.~DMR-0213745 and by the Institute for
Theoretical Sciences
and DOE, No.  W-31-109-ENG-38 (QC).

\bibliographystyle{apsrev}

\begin{thebibliography}{29}
\bibitem[{foo()}]{footnoteonN0}
\bibinfo{note}{Defining the condensate fraction in the fermionic regime is not
  unambiguous. $N_s/N$ which we use here gives the largest possible fraction.}
\end{thebibliography}

\end{document}